**Average causal effect estimation via instrumental variables: the no simultaneous heterogeneity assumption**

Fernando Pires Hartwig[1,2]*, Linbo Wang[3], George Davey Smith[2,4], Neil Martin Davies[2,4,5]

[1]Postgraduate Program in Epidemiology, Federal University of Pelotas, Pelotas, Brazil.

[2]MRC Integrative Epidemiology Unit, University of Bristol, Bristol, United Kingdom.

[3]Department of Statistical Sciences, University of Toronto, Toronto, ON, Canada.

[4]Population Health Sciences, Bristol Medical School, University of Bristol, Bristol, United Kingdom.

[5]K.G. Jebsen Center for Genetic Epidemiology, Department of Public Health and Nursing, NTNU, Norwegian University of Science and Technology, Norway.

*Corresponding author. Postgraduate Program in Epidemiology, Federal University of Pelotas, Pelotas (Brazil). Adress: Rua Marechal Deodoro 1160 (3rd floor). ZIP code: 96020-220. Phone: 55 53 981126807. E-mail: fernandophartwig@gmail.com.

**Abstract**

**Background:** Instrumental variables (IVs) can be used to provide evidence as to whether a treatment $X$ has a causal effect on an outcome $Y$. Even if the instrument $Z$ satisfies the three core IV assumptions of relevance, independence and the exclusion restriction, further assumptions are required to identify the average causal effect (ACE) of $X$ on $Y$. Sufficient assumptions for this include: homogeneity in the causal effect of $X$ on $Y$; homogeneity in the association of $Z$ with $X$; and no effect modification (NEM).

**Methods:** We describe the NO Simultaneous Heterogeneity (NOSH) assumption, which requires the heterogeneity in the $X$-$Y$ causal effect to be mean independent of (i.e., uncorrelated with) both $Z$ and heterogeneity in the $Z$-$X$ association. This happens, for example, if there are no common modifiers of the $X$-$Y$ effect and the $Z$-$X$ association, and the $X$-$Y$ effect is additive linear. We illustrate NOSH using simulations and by re-examining selected published studies.

**Results:** When NOSH holds, the Wald estimand equals the ACE even if both homogeneity assumptions and NEM (which we demonstrate to be special cases of – and therefore stronger than – NOSH) are violated.

**Conclusions:** NOSH is sufficient for identifying the ACE using IVs. Since NOSH is weaker than existing assumptions for ACE identification, doing so may be more plausible than previously anticipated.

## 1. Introduction

Instrumental variable (IV) methods can be applied to strengthen causal inference using non-experimental data when the assumption of no unmeasured confounding is implausible.[1–3] A valid IV satisfies the three core assumptions (defined formally below) of relevance, independence and the exclusion restriction.[4,5] These assumptions allow testing the sharp causal null hypothesis – i.e., whether the treatment has a causal effect on the outcome for at least one member of the population. However, point-estimating well-defined treatment effect parameters requires further assumptions, sometimes referred to as fourth point identifying assumptions (IV4).[6] In this paper our target estimand is the average causal effect ($\mathrm{ACE}$) on the additive scale (defined precisely in section 2.1). Different IV4 assumptions sufficient for ACE identification have been proposed, including: treatment effect homogeneity,[4,7] instrument effect homogeneity,[8,9] no effect modification (NEM)[4,9], and no unmeasured common effect modifier (NUCEM)[10] (defined precisely in section 2.3). Given our focus on the ACE, IV4 assumptions that identify other estimands (e.g., monotononicity, which identifies the local average causal effect (LACE)[11]) will not be discussed in detail.

Previous papers have described the IV estimand when treatment effects vary. Heckman (1997) described, for a binary treatment, that if individual treatment effects do not affect whether someone is treated, the IV estimand equals the $\mathrm{ACE}$.[7] Harris and Remler (1998) noted this, stating that the ACE is identified if treatment effect heterogeneity is unrelated to treatment assignment.[12] Brookhart and Schneeweiss (2007) described how, if the individual levels effects of a binary treatment are the same for different instrument values, the IV estimator identifies the $\mathrm{ACE}$ in the population.[8] Wang and Tchetgen Tchetgen (2018) describe how, for a binary instrument and a binary treatment, the $\mathrm{ACE}$ is identified if no unmeasured confounders are additive modifiers of the association between the instrument and the treatment or of the effect of the treatment on the outcome.[9] Cui and Tchetgen Tchetgen (2021), again focusing on binary instrument and treatment, proposed a weaker version of this condition which only requires that there is no common additive modifier of the instrument-treatment association and the treatment-outcome effect.[10] Syrgkanis et al. (2019) state that the $\mathrm{ACE}$ is not generally identified if individual-level instrument-treatment and treatment-outcome effects are dependent.[13]

Numerous reviews and methodological papers have described several identifying assumptions.[4,14–17] Here, we introduce the NO Simultaneous Heterogeneity (NOSH) assumption. We show that if NOSH holds, the IV estimand equals the $\mathrm{ACE}$. We also show that other IV4 assumptions are special cases of NOSH. Finally, we use simulations to corroborate the theory.

## 2. Methods

### 2.1. Notation and assumptions

Let $Z$, $X$, $Y$ and $U$ respectively denote the instrument, the time-fixed treatment, the outcome and all unmeasured common causes of $X$ and $Y$. For simplicity, we consider the case where no adjustment is made for measured covariates. However, the concepts developed here can be trivially extended to accommodate measured covariates. We also discuss covariate adjustment in the simulation study (section 4 in the Supplement).

$Z$ is a valid IV if the following three causal assumptions (illustrated in Figure 1) are satisfied: i) relevance: $Z$ and $X$ are statistically dependent; ii) independence: $Z \perp\!\!\!\perp U$ (where "$\perp\!\!\!\perp$" denotes statistical independence); and iii) exclusion restriction: $Z \perp\!\!\!\perp Y|X, U$.[4,5]

Figure 1 is a graphical representation of the following non-parametric structural equation model:

$$X_i = f_X\left(Z = Z_i, U = U_i, \varepsilon_X = \varepsilon_{X_i}\right)$$

$$Y_i = f_Y\left(X = X_i, U = U_i, \varepsilon_Y = \varepsilon_{Y_i}\right),$$

where $X_i$ is the value of $X$ for individual $i$ (the same notation applies to other variables), $f_X$ and $f_Y$ respectively denote the functions governing $X$ and $Y$, and $\varepsilon_X$ and $\varepsilon_Y$ respectively denote stochastic direct causes of $X$ and $Y$ (so that $\varepsilon_X \perp\!\!\!\perp \varepsilon_Y$). We now define $F_{X_i}(z) = \mathrm{E}\left[X_i | do(Z = z), U = U_i, \varepsilon_X = \varepsilon_{X_i}\right]$ and $F_{Y_i}(x) = \mathrm{E}\left[Y_i | do(X = x), U = U_i, \varepsilon_X = \varepsilon_{X_i}\right]$ – that is, $F_{X_i}(z)$ is the expectation of $X$ when $Z$

is set (possibly counterfactually) to $z$, while the other variables retain their observed values. A similar interpretation holds for $F_{Y_i}(x)$.

The individual level instrumental and treatment effects $\beta_X$ and $\beta_Y$ can be defined as follows:

$\beta_{X_i} = F_{X_i}(1) - F_{X_i}(0)$ for a binary $Z$,

$\beta_{X_i} = F_{X_i}(z) - F_{X_i}(z-1)$ for a multivalued discrete $Z$,

$\beta_{X_i} = \frac{\partial}{\partial z}\left[F_{X_i}(z)\right]\Big|_{z=Z_i}$ for a continuous $Z$,

$\beta_{Y_i} = \mathrm{E}\left[F_{Y_i}(1) - F_{Y_i}(0) | U = U_i, \varepsilon_Y = \varepsilon_{Y_i}\right]$ for a binary $X$,

$\beta_{Y_i} = \frac{\partial}{\partial x}\left[F_{Y_i}(x)\right]\Big|_{x=X_i}$ for a continuous $X$.

For a multivalued discrete $Z$, we assume $Z$ is coded numerically such that $\mathrm{E}[X|Z=1] \leq \cdots \leq \mathrm{E}[X|Z=K]$, where $K$ is the number of values that $Z$ attains, and $z \in \{2, \ldots, K\}$. Notice that, for continuous $Z$, the definition of $\beta_X$ implicitly assumes that $F_X(z)$ is differentiable with respect to $Z$. For non-continuous $X$, this would happen for example if $X_i \sim \mathrm{Bernoulli}(p_i)$ or $X_i \sim \mathrm{Poisson}(\lambda_i)$, where $p_i$ or $\lambda_i$ are differentiable functions of $Z$. A similar notion applies to $\beta_Y$ and $F_Y(z)$ for a continuous $X$.

Under the stable unit treatment value assumption, $\beta_{X_i}$ is the additive change in the expectation of $X$ caused by a unit increase in $Z$ in individual $i$ and $\beta_{Y_i}$ is the additive change in the expectation of $Y$ caused by a unit increase in $X$ in individual $i$. From this notation, the ACE is defined as $\mathrm{E}[\beta_Y]$ – i.e., the average of $\beta_Y$ in the population. This definition incorporates the case of a multi-valued $X$ (excluding the case where $X$ is an unordered multivalued variable or $X$ is a discrete variable with non-linear effects on $Y$, since derivatives are not defined in these cases), where the distribution of $X$ in the population will affect the ACE. This quantity is sometimes referred to as the average derivative effect.[18,19] Finally, the conventional IV estimand known as the Wald estimand (here denoted as $\beta_{IV}$) is defined as $\beta_{IV} = \frac{\mathrm{cov}(Y,Z)}{\mathrm{cov}(X,Z)}$.[13,20]

## 2.2. The NO Simultaneous Heterogeneity (NOSH) assumption

We define the NOSH assumption as a combination of two conditions (Assumptions 1 and 2, defined below). The name NOSH refers to the fact that, if these two assumptions hold, then $\beta_Y \perp\!\!\!\perp (Z, \beta_X)$ – that is, heterogeneity in the causal effect is independent of heterogeneity in the instrument effect (and of the instrument). Since $\beta_X$ and $\beta_Y$ denote effects on the additive scale, NOSH focuses on additive effect modification. Of note, there is no assumption regarding multiplicative effect modification other than what is implied by our assumptions on additive effect modification (in section 2 in the supplement, we discuss implications of non-linear effects and non-linear data-generating models for NOSH).

**Theorem 1**: If NOSH holds, then $\beta_{IV} = \mathrm{ACE}$ (proof in section 3 in the Supplement).

We now define the conditions for NOSH to hold using causal diagrams (see section 1 in the Supplement for equivalent definitions using non-parametric structural equation models). For a precise articulation, it is useful to partition $U$ in Figure 1 in six non-overlapping, exhaustive sets of variables (Table 1).

Figure 2A illustrates possible causal relationships compatible with Figure 1 (i.e., compatible with the core IV assumptions) among $U$, $Z$, $X$, $\beta_X$ and $\beta_Y$. Of note, Figure 1 assumes that $Z$ is a causal instrument, but this is not necessary for NOSH to be defined or hold. In Theorem 1, $\beta_X$ can be replaced $\beta_X^*$, here denoting the individual-level association of a non-causal instrument $Z^*$ and $X$. By non-causal instrument, we mean that $Z^*$ is not a cause of $X$. Still, it is associated with $X$ through paths that included $Z$ as a non-collider (more specifically, paths of the form $Z^* \leftarrow W \rightarrow Z$ or $Z^* \rightarrow \boxed{C} \leftarrow Z$, where $C$ is being conditioned on). Therefore, both modifiers of the effect of $Z$ on $X$ and modifiers of the association between $Z^*$ and $Z$ will be modifiers of the association of $Z^*$ and $X$. Modifiers of $Z$-$X$ must be independent of modifiers of $X$-$Y$ for NOSH to hold, as discussed above. Moreover, modifiers of $Z$-$Z^*$ must be independent of any cause of $Y$ (otherwise $Z^*$ would be an invalid IV). Therefore, if Assumptions 1 and 2 (see below) hold for $Z$, they also hold for $Z^*$. Given this clarification, $Z$ will be depicted as a causal instrument in Figure 2 for simplicity.

Although it is not usual to depict individual-level effects as nodes in a causal graph, $\beta_X$ and $\beta_Y$ are indeed random variables. Since $X$ is fixed in time, these two variables are simply functions of other individual-level random variables, so they are not qualitatively different from $X$ or $Y$ for example. This contrasts with variables that may also depend on non-individual level characteristics, such as person-time, which depends on follow-up duration and can be influenced by study design. Causes of $\beta_X$ can be interpreted as modifiers of the effect of $Z$ on $X$, while causes of $\beta_Y$ can be interpreted as modifiers of the effect of $X$ on $Y$. A more comprehensive description on representing individual-level effects in causal graphs is available elsewhere.[21] For example, Table 1 describes that $U_2$ modifies the effect of $X$ on $Y$, but not the effect of $Z$ on $X$. This is translated in Figure 2A by the directed path from $U_2$ to $\beta_Y$, but not to $\beta_X$ (see Box 1). Explicitly depicting $\beta_X$ and $\beta_Y$ in the graph facilitates identifying conditions under which $\beta_X$ and $\beta_Y$ are d-separated, which is useful to precisely articulate sufficient conditions for NOSH to hold.

In Figure 2A, all unmeasured variables $U_1$ to $U_6$ are d-connected to one another due to a latent variable, thus allowing for statistical dependencies between them in unrestricted ways. Although these variables could be d-connected due to other causal structures (e.g., a path $U_1 \rightarrow U_2$), this would violate the classification proposed in Table 1 (in this example, $U_1$ would be an effect modifier of both the effect of $Z$ on $X$ and the effect of $X$ on $Y$ – i.e., it would be a component of $U_6$), and thus blur the distinct implications of distinct types of effect modifiers.

Although this is not central to our arguments, it is instructive to clarify that, from a statistical perspective, both $U_3$ and $U_4$ (for example) are effect modifiers of $\beta_X$ if $U_3$ and $U_4$ are d-connected, because $\beta_X$ will vary between strata of $U_3$ and between strata of $U_4$ (the same reasoning applies to other unmeasured variables with respect to $\beta_X$ and/or $\beta_Y$). For clarity, we use “effect modifier” to refer to variables that themselves exert effect modification and “surrogate effect modifier” to refer to variables that are d-connected with effect modifiers but do not themselves exert effect modification. Although our arguments could ignore surrogate effect modifiers, defining different types of effect modifiers allows for a more comprehensive articulation of conditions sufficient for NOSH to hold.

Since d-separation implies statistical independence, causal diagrams can be used to find conditions under which $\beta_Y$ is d-separated from (and thus statistically independent of) $\beta_X$ and $Z$. Under such conditions, NOSH holds. In Figure 2A, $\beta_Y$ is d-connected through multiple paths to both $Z$ and $\beta_X$, so $\beta_Y \perp\!\!\!\perp (Z, \beta_X)$ (and therefore NOSH) will not generally hold.

We now describe the assumptions that define NOSH.

**Assumption 1**: All unmeasured variables that modify $Z$-$X$ are independent of all unmeasured variables that modify $X$-$Y$.

This assumption implies that there are no unmeasured variables that themselves modify (on the additive scale) both the effect of $Z$ on $X$ and the effect of $X$ on $Y$ – that is, $U_6 = \emptyset$. Furthermore, all unmeasured variables that modify the effect of $Z$ on $X$ (i.e., $U_1$ and $U_4$) are independent of all unmeasured variables that modify the effect of $X$ on $Y$ (i.e., $U_2$ and $U_5$). This implies that there are neither unmeasured effect modifiers of $Z$-$X$ that are also surrogate effect modifiers of $X$-$Y$ nor unmeasured surrogate effect modifiers of $Z$-$X$ that are also effect modifiers of $X$-$Y$. However, $U_1$, $U_3$ and $U_4$ may be correlated with one another, and $U_2$, $U_3$ and $U_5$ may be correlated with one another. For this to hold, $X$ cannot cause $U_2$. Otherwise, $U_1$ and $U_4$ (which are modifiers of the effect of $Z$ on $X$) and $U_2$ (which is a modifier of the effect of $X$ on $Y$) will be d-connected through the paths $U_1 \rightarrow X \rightarrow U_2$ and $U_4 \rightarrow X \rightarrow U_2$. This assumption is violated in Figure 2A, where all unmeasured variables are allowed to be d-connected with one another.

Even if Assumption 1 holds, paths of the form $\beta_X \leftarrow Z \rightarrow X \rightarrow \beta_Y$, which render $\beta_X$ and $\beta_Y$ d-connected, may still exist. Therefore, Assumption 1 is necessary, but not sufficient, to render $\beta_X$ and $\beta_Y$ d-separated.

**Assumption 2**: The effect of $X$ on the expectation of $Y$ is additive linear.

This assumption holds if $F_{Y_i}(x) = \varphi_i + \mu_i x$, where $\varphi_i$ and $\mu_i$ may vary between individuals. In this case, $F_{Y_i}(x) - F_{Y_i}(x') = \mu_i(x - x')$. This implies that, for a given individual, the additive change in the expectation of $Y$ caused by a unit increase in $X$ does not depend on the value of $X$ that the individual has. That is, the effect of $X$ on $Y$ is additive linear (but not necessarily constant) across individuals. This implies that the path $X \rightarrow \beta_Y$ does not exist. Consequently, $Z$ and $\beta_Y$ are d-separated (that is, $\beta_Y \perp\!\!\!\perp Z$). Of note, this assumption is automatically satisfied if $X$ is binary.

In Figure 2B, both Assumptions 1 and 2 hold. In this graph, $\beta_Y$ and $\beta_X$ are d-separated (because all paths from $\beta_X$ to $\beta_Y$ contain at least one collider), and $\beta_Y$ and $Z$ are d-separated (because all paths from $Z$ to $\beta_Y$ contain $X$ as a collider).

Even though the above focused on the structural interpretation of NOSH (i.e., an interpretation where concepts can be represented in causal graphs), this assumption can be relaxed in the sense that it does not require full independence, but only mean independence (i.e., uncorrelatedness). That is, if $\mathrm{E}[\beta_Y | Z, \beta_X] = \mathrm{E}[\beta_Y]$, then $\beta_{IV} = \mathrm{ACE}$ (see the proof in the Supplement for details). Therefore, NOSH is a statistical statement, so representing it using causal graphs may be useful, but not strictly required. For example, In Figure 2A, even though NOSH will not generally hold, it is possible that it holds under lack of faithfulness (i.e., when there is d-connection but no statistical association). In this sense, NOSH is agnostic to whether faithfulness is assumed.

## 2.3. NOSH is a generalization of previous IV4 assumptions

We now show that well-known assumptions that identify the $\mathrm{ACE}$ imply NOSH.

### 2.3.1. Causal effect homogeneity

For a binary $X$ and assuming deterministic counterfactuals, this assumption can be defined as $Y_i(X_i = 1) - Y_i(X_i = 0) = c$ (a constant), where $Y_i(X_i = x) = f_Y\big(do(X = x), U = U_i, \varepsilon_Y = \varepsilon_{Y_i}\big)$ for $x \in \{0,1\}$. More generally, this condition can be defined as $\beta_{Y_i} = c$. Since $\beta_Y$ is constant, NOSH trivially holds – i.e., NOSH is implied by causal effect homogeneity. Moreover, NOSH allows identification when there is causal effect heterogeneity. Therefore, NOSH is weaker than causal effect homogeneity.

### 2.3.2. Instrument effect homogeneity

For a binary causal instrument $Z$ and assuming deterministic counterfactuals, this assumption can be defined as $X_i(Z_i = 1) - X_i(Z_i = 0) = c$, where $X_i(Z_i = Z) = f_X\big(do(Z = z), U = U_i, \varepsilon_X = \varepsilon_{X_i}\big)$ for $x \in \{0,1\}$. Generally, this condition can be defined as $\beta_{X_i} = c$. Since $\beta_X$ is constant, $\beta_Y \perp\!\!\!\perp \beta_X$ trivially holds. However, instrument effect homogeneity does not imply $\beta_Y \perp\!\!\!\perp Z$, because the effect of $X$ on $Y$ may be non-linear, except if $X$ is binary. This is important because additive linearity in the effect of $X$ on $Y$ is required for the conventional IV estimand to equal the $\mathrm{ACE}$[22] (see section 3 in the Supplement).

Therefore, instrument effect homogeneity implies that Assumption 1 is true. However, $\mathrm{ACE}$ identification also requires Assumption 2, which means NOSH holds. Since NOSH allows identification under instrument effect heterogeneity and both NOSH and instrument effect homogeneity require Assumption 2, NOSH is weaker than instrument effect homogeneity.

### 2.3.3. No effect modification (NEM) by unmeasured factors

Homogeneity can be relaxed by considering a condition sometimes referred to as NEM. We consider two versions of this assumption: NEM1 and NEM2. For a binary $X$, NEM1 is defined as $\mathrm{E}[Y_i(X_i = 1) - Y_i(X_i = 0)|U_i] = \mathrm{E}[Y_i(X_i = 1) - Y_i(X_i = 0)]$.[9] More generally, NEM1 postulates that $\mathrm{E}\big[\beta_{Y_i}|U_i\big] = \mathrm{E}\big[\beta_{Y_i}\big]$ – that is, no unmeasured $X$-$Y$ confounder modifies the additive effect of $X$ on the expectation of $Y$. For a binary $X$, NEM1 holds if $U$ and $\beta_Y$ are d-separated; otherwise, it would not hold in general. Therefore, Assumption 1 is necessary for NEM1 to hold; otherwise, there will be open paths between $\beta_Y$ and (components of) $U$. However, it is not sufficient since Assumption 1 allows for confounders to be effect modifiers. If $X$ is continuous, Assumption 2 is also necessary (but not sufficient), otherwise the path $X \rightarrow \beta_Y$ will exist, which would render $\beta_Y$ and $U$ d-connected, thus violating NEM1. Since Assumptions 1 and 2 are both necessary for NEM1 and sufficient for NOSH to hold, NEM1 implies NOSH. However, since these two assumptions are insufficient for NEM1 to hold, NOSH does not imply NEM1. Therefore, NOSH is weaker than NEM1. Of note, some authors refer to violation of NEM1 as essential heterogeneity.[23]

Although not typically referred to this way, NEM also applies to the association between $Z$ and $X$ (we will call this condition NEM2). For a binary causal instrument $Z$, NEM2 is defined as $\mathrm{E}[X_i(Z_i = 1) - X_i(Z_i = 0)|U_i] = \mathrm{E}[X_i(Z_i = 1) - X_i(Z_i = 0)]$.[9] More generally, NEM2 postulates that $\mathrm{E}\left[\beta_{X_i}|U_i\right] = \mathrm{E}\left[\beta_{X_i}\right]$ – that is, no unmeasured $X$-$Y$ confounder modifies the additive association between $Z$ and $X$. If both $Z$ and $X$ are binary, NEM2 holds if $U$ and $\beta_X$ are d-separated; otherwise, it would not hold in general. Therefore, Assumption 1 is necessary for NEM2 to hold; otherwise, there will be open paths between $\beta_Y$ and (components of) $U$. However, it is not sufficient, since Assumption 1 allows for confounders to be effect modifiers. For a continuous $X$, NEM2 is not sufficient to identify the $\mathrm{ACE}$ since even the stronger condition of instrument effect homogeneity requires Assumption 2. Since Assumptions 1 and 2 are both necessary for NEM2 to identify the $\mathrm{ACE}$ and sufficient for NOSH to hold, NEM2 implies NOSH. However, since these two assumptions are insufficient for NEM2 to hold, NOSH does not imply NEM2. Therefore, NOSH is weaker than NEM2.

#### 2.3.4. No unmeasured effect modification (NUCEM)

For binary $Z$ and $X$, it has been shown that the usual IV estimand equals the ACE if $\mathrm{Cov}(\mathrm{E}[X|U, Z = 1] - \mathrm{E}[X|U, Z = 0], \mathrm{E}[Y|U, do(X = 1)] - \mathrm{E}[Y|U, do(X = 1)]) = 0$. In words, this condition (which is clearly weaker than NEM), postulates that there are no unmeasured variables that modify (on the additive scale) both the $Z$-$X$ association and the $X$-$Y$ effect.[10] This zero-covariance condition can be expressed as $\mathrm{E}[\beta_Y|\beta_X] = \mathrm{E}[\beta_Y]$, which is equivalent to NOSH for binary $Z$ and $X$ (this is because, for a binary $X$, $\mathrm{E}[\beta_Y|Z, \beta_X] = \mathrm{E}[\beta_Y|\beta_X]$ since in this case Assumption 2 is guaranteed to hold). Therefore, NOSH generalizes NUCEM to situations involving non-binary $Z$ and $X$.

## 3. Simulation study

We performed a simulation to demonstrate further NOSH is sufficient for ACE identification (a detailed description is provided in section 4 in the Supplement). Results are shown in Figure 3. In scenario 1, NOSH holds, and the two-stage least squares estimator (TSLS) (equivalent to the Wald estimator for a single $Z$, $X$ and $Y$[24,25]) consistently estimates the $\mathrm{ACE}$, with coverage being approximately 95% and bias converging to zero as sample size increases. A similar pattern was seen when NOSH holds, but error terms were non-normal (scenarios 4 and 5). When NOSH is violated (scenarios 2 and 3), TSLS had substantial bias and undercoverage. Supplementary Table 1 shows the results for different TSLS specifications. When NOSH is violated, but Assumption 2 holds (as in scenario 2), adjusting for measured common effect modifiers mitigates bias. However, when NOSH is violated exclusively due to Assumption 2 being invalid (as in scenario 3), covariate adjustment does not improve estimation.

## 4. Discussion

This paper shows that the Wald estimator is consistent for the $\mathrm{ACE}$ if, in addition to the core IV assumptions, the NOSH assumption holds. NOSH is weaker than previously proposed IV4 assumptions that identify the $\mathrm{ACE}$. This does not include the monotonicity assumption (defined precisely in section 5.1 in the Supplement), which is sometimes classified as an IV4 assumption but does not identify the ACE. While NOSH is not strictly weaker or stronger than monotonicity, NOSH has two advantages over monotonicity. First, the latter only identifies $\mathrm{LACE}$s, whereas NOSH identifies the $\mathrm{ACE}$. Second, the NOSH assumption identifies the $\mathrm{ACE}$ even if there are defiers. However, if NOSH is violated but monotonicity holds, IV estimators, identify the $\mathrm{LACE}$, which is a well-defined causal parameter for binary treatments. However, the notion of compliers is not well-defined for continuous treatments. In this case, monotonicity allows interpreting the Wald estimand as a weighted average of treatment effects, with subgroups of the population where the $Z$-$X$ association is stronger receiving greater weight.[20,26] Although mathematically well-defined, this parameter is difficult to interpret for policy making. Conversely, if NOSH holds, then IV estimators will identify the $\mathrm{ACE}$ of a continuous treatment.

In many recent papers describing methodological developments that relax the exclusion restriction assumption when there are multiple IVs, treatment effect homogeneity is implicitly or explicitly required.[27,28] This is because these methods assume that valid IVs identify the same causal effect (generally, the $\mathrm{ACE}$). However, assuming NOSH holds for all IVs is sufficient to identify the $\mathrm{ACE}$. This implies that the assumptions required for the validity of these methods are weaker than previously considered. Nevertheless, since we defined NOSH for a time-fixed treatment, caution must be taken to extrapolate our conclusions to time-varying treatments. Moreover, even though NOSH is weaker than homogeneity or NEM, it is still quite strong and should not be taken for granted in practice.

NOSH is an untestable assumption that cannot be guaranteed by study design. Therefore, assessing its plausibility requires subject matter knowledge. We illustrate this by discussing three published IV studies (section 5 in the Supplement). The possibility of (partially) empirically verifying some IV4 assumptions has implications for NOSH plausibility. For example, in the case of a continuous treatment, instrument effect heterogeneity would often imply (except in some specific circumstances) that the treatment is heteroscedastic with respect to the instrument (i.e., the variance of the treatment would differ between levels of the instrument).[29,30] Future methodological studies are required to assess the power and utility of such tests in typical IV settings. Since NOSH is weaker than homogeneity, homoscedasticity, in this case, would support (but not guarantee) that NOSH holds. However, heteroscedasticity would not necessarily imply that NOSH is violated. Brookhart (2007)[8] proposed empirically assessing the plausibility of instrument effect homogeneity by testing if instrument strength varies between strata of measured covariates. This can also be viewed as a test of NOSH since if multiple covariates modify instrument strength, then the assumption that there are no unmeasured common effect modifiers is less plausible. Indeed, assuming $Z$ is a valid instrument, variability in IV estimates between strata defined by such covariates could be interpreted as evidence against NOSH. Strategies to empirically verify this assumption remain to be formally investigated.

An earlier glimpse of NOSH in the literature can be found, for example, in Angrist (1990),[31] who noted that variation in response to the draft (i.e., $Z$-$X$ heterogeneity) based on potential outcomes (i.e., $X$-$Y$ heterogeneity) could mean his results were biased estimates of the $\mathrm{ACE}$. Indeed, this would be a scenario where NOSH is violated. For binary $Z$ and $X$, NOSH is equivalent to NUCEM.[10] However, no causal structures (e.g., in the form of causal graphs) that dictate whether NOSH is violated were presented. Syrgkanis et al. (2019)[13] also postulated an independence condition that is equivalent to NOSH, without necessarily restricting to binary $Z$ and $X$. However, mechanisms that could render this condition satisfied were similarly not discussed, and no explicit consideration was given to potential implications of non-linear instrument-outcome associations or treatment-outcome effects. Moreover, their assumed data-generating mechanism for the outcome was $Y_i = \theta_i(U = U_i)X_i + \rho(U = U_i) + e_i$, where $\theta(U)$ is the causal effect function, which was assumed to be linear additive. However, such a model may not be appropriate when $Y$ is binary, which is often assumed to be a non-linear function of $X$ and $U$ (e.g., the expectation of $Y$ may be a logistic function of $X$ and $U$).

Here we use counterfactual notation to explicitly define individual-level effects in a framework that allows for both binary and continuous $Z$, $X$ and/or $Y$. We also comprehensively describe data-generating mechanisms influencing NOSH using causal graphs. This helps to apply expert knowledge to assess the plausibility of this assumption in practice. We also explicitly consider the implications of non-linear $Z$-$X$ associations and $X$-$Y$ effects. The present work thus clarifies the assumptions underlying NOSH, which allows differentiating it from previous IV4 assumptions and explicitly propose NOSH as an IV4 assumption that is weaker than previously described IV4 assumptions that identify the $\mathrm{ACE}$.

## 5. Acknowledgments

The Medical Research Council (MRC) and the University of Bristol support the MRC Integrative Epidemiology Unit [MC_UU_00011/1]. NMD is supported by an Economics and Social Research Council (ESRC) Future Research Leaders grant [ES/N000757/1] and a Norwegian Research Council

Grant number 295989. LW is partially supported by a McLaughlin Accelerator Grant in Genomic Medicine.

This work is dedicated to the memory of Mr. Dari Hartwig (FPH's father).

**Table 1. Subsets of unmeasured variables collectively represented as $U$ in Figure 1.**

| Variable | Causes $X$ | Causes $Y$ | Modifies* $Z$- $X$ | Modifies* $X$- $Y$ |
|---|---|---|---|---|
| $U_1$† | Yes | No | Yes | No |
| $U_2$† | No | Yes | No | Yes |
| $U_3$ | Yes | Yes | No | No |
| $U_4$ | Yes | Yes | Yes | No |
| $U_5$ | Yes | Yes | No | Yes |
| $U_6$ | Yes | Yes | Yes | Yes |

†Strictly speaking, $U_1$ and $U_2$ do not belong to $U$ in Figure 1, but we will use the notation $U_j$ to refer to unmeasured variables in general.

*This refers to additive effect modification.

**Figure 1. Causal graph illustrating the instrumental variable assumptions.**

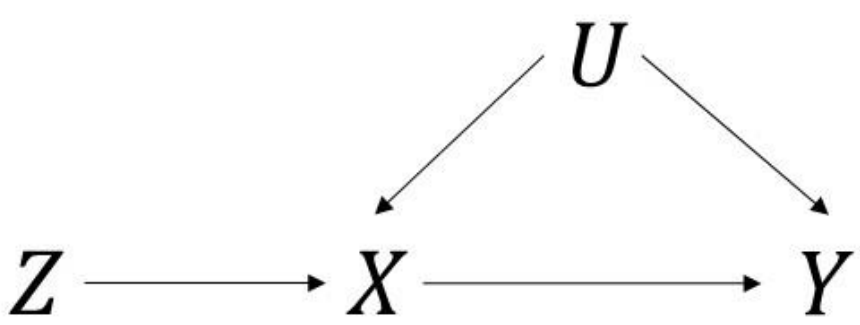

**Figure 2. Causal graph illustrating causal relationships between the instrument ($Z$), the treatment ($X$), individual level causal effect of $Z$ on $X$ ($\beta_X$)*, individual level causal effect of $X$ on the $Y$ ($\beta_Y$)* and six unmeasured variables ($U_1$ to $U_6$). The dotted lines represent causal effects from unknown causes ($L$, $L_1$ and $L_2$, with a dotted box marking them as latent) that render some or all unmeasured variables d-connected.**

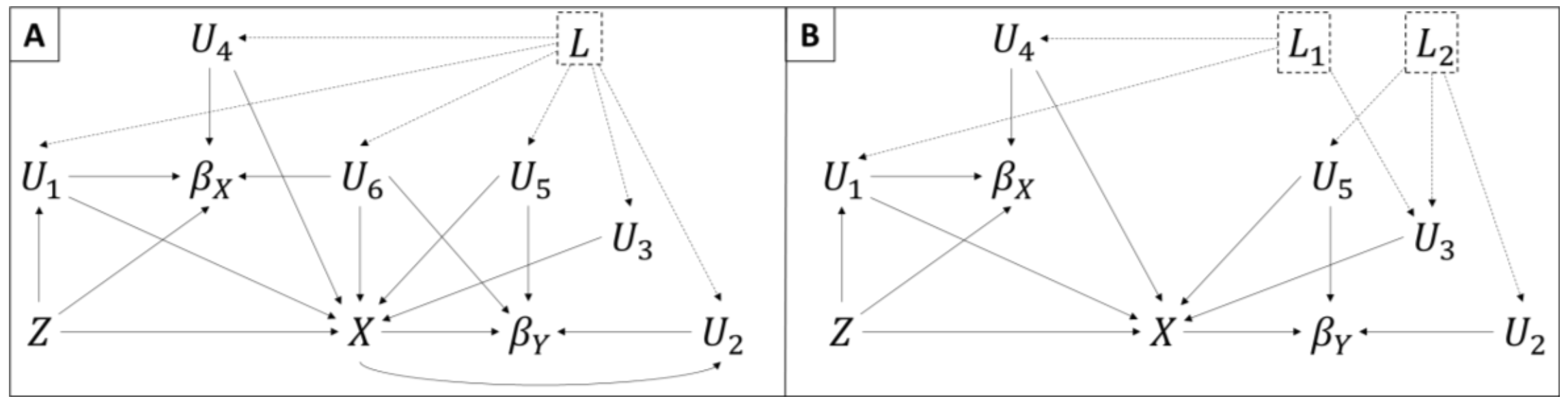

A: All possible causal relationships compatible with Figure 1 (i.e., unrestricted model, where assumptions 1 and 2 are violated).

B: Assumptions 1 and 2 hold.

*Since $\beta_X$ and $\beta_Y$ represent effects on the additive scale, variables d-connected with them are additive effect modifiers of $Z$-$X$ and $X$-$Y$, respectively.

**Figure 3. Median bias, median standard error, coverage and rejection rate of two-stage least squares as an estimator of the ACE† in scenarios 1-5‡, where the causal effect of $X$ on $Y$ is linear.**

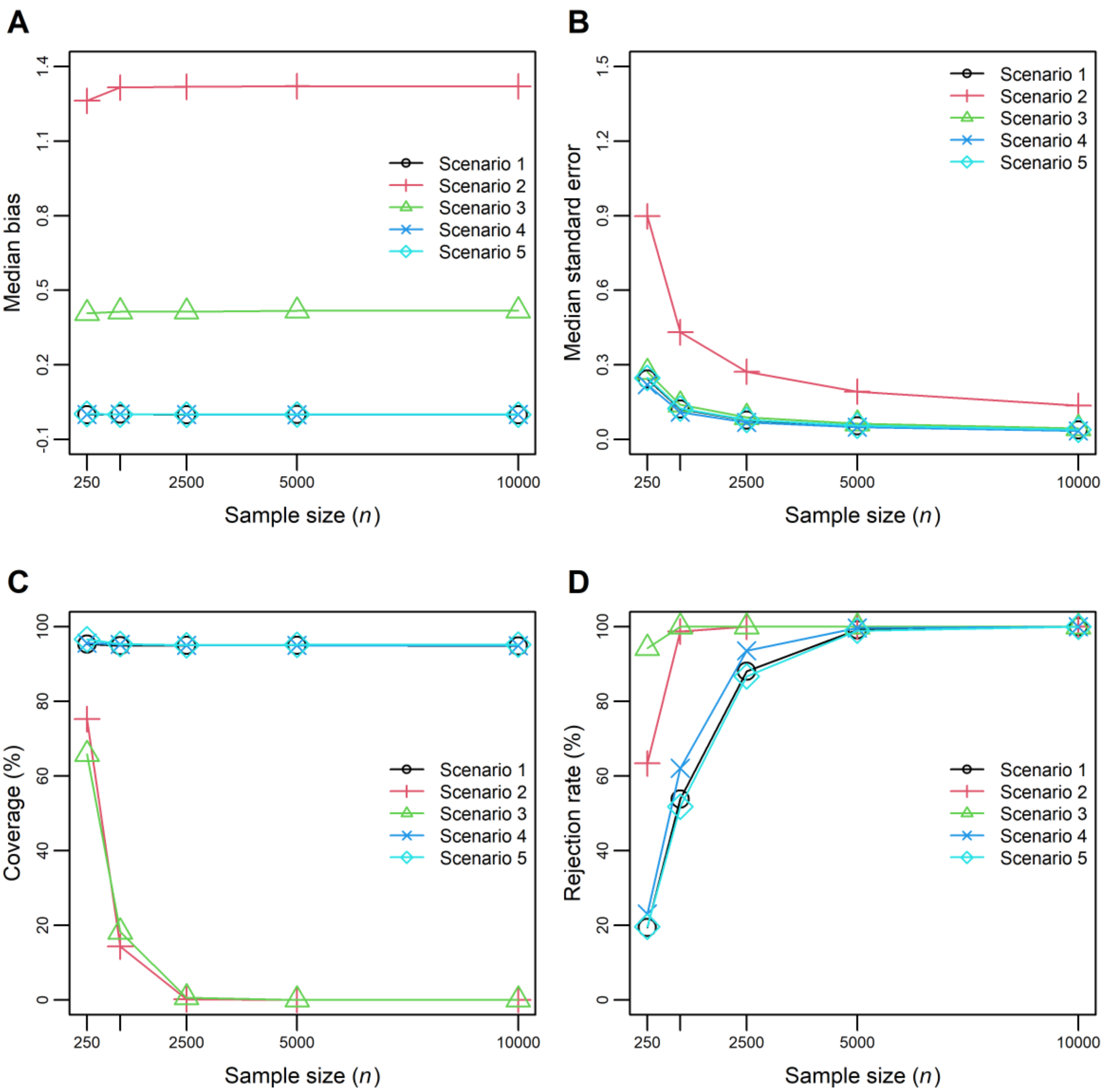

†Average causal effect (ACE).

‡1: NOSH holds. 2: Assumption 1 violated. 3: Assumption 2 violated. 4: NOSH holds and error terms are drawn from a beta distribution. 5: NOSH holds and error terms are drawn from a mixed chi-squared distribution.

**Box 1. Illustration of the equivalence between effect modification and causes of individual level effects**

Graphically representing effect modification as causes of individual level effects is consistent with the definition of effect modification. To illustrate, suppose $Y_i = \alpha_1 + \alpha_2 X_i + \alpha_3 X_i V_i + \varepsilon_i$, where $V_i$ is not caused by $X$ and $\varepsilon_i$ is some continuous random error. In this situation, $\beta_Y$ is defined as follows:

$$\beta_{Y_i} = \frac{\partial}{\partial X}\left[\beta_{Y_i}\right] = \alpha_2 + \alpha_3 V_i.$$

In this example, $V$ is a cause of $\beta_Y$ because it is in the right-hand side of the structural equation model for $\beta_Y$. This would be represented in a causal graph as any other causal relationship – i.e., as a directed path from the cause ($V$) to the consequence ($\beta_Y$).

**Supplementary material for "Average causal effect estimation via instrumental variables: the no simultaneous heterogeneity assumption" by Hartwig FP, Wang L, Davey Smith G and Davies NM**

**1. Non-parametric structural equation models (SEMs)**

In this section we describe the SEMs corresponding to the causal graphs and assumptions required for NOSH to hold. Initially, the SEM corresponding to the unrestricted model (depicted in Figure 2A) is:

$$U_{1i} = f_{U_1}\left(Z = Z_i, \varepsilon_L = \varepsilon_{L_i}, \varepsilon_{U_1} = \varepsilon_{U_{1i}}\right) \quad \text{(S1.1)}$$

$$U_{2i} = f_{U_2}\left(X = X_i, \varepsilon_L = \varepsilon_{L_i}, \varepsilon_{U_2} = \varepsilon_{U_{2i}}\right) \quad \text{(S2.1)}$$

$$U_{3i} = f_{U_3}\left(\varepsilon_L = \varepsilon_{L_i}, \varepsilon_{U_3} = \varepsilon_{U_{3i}}\right) \quad \text{(S3.1)}$$

$$U_{4i} = f_{U_4}\left(\varepsilon_L = \varepsilon_{L_i}, \varepsilon_{U_4} = \varepsilon_{U_{4i}}\right) \quad \text{(S4.1)}$$

$$U_{5i} = f_{U_5}\left(\varepsilon_L = \varepsilon_{L_i}, \varepsilon_{U_5} = \varepsilon_{U_{5i}}\right) \quad \text{(S5.1)}$$

$$U_{6i} = f_{U_6}\left(\varepsilon_L = \varepsilon_{L_i}, \varepsilon_{U_6} = \varepsilon_{U_{6i}}\right) \quad \text{(S6.1)}$$

$$Z_i = f_Z\left(\varepsilon_Z = \varepsilon_{Z_i}\right) \quad \text{(S7.1)}$$

$$X_i = f_X\left(Z = Z_i, U_1 = U_{1i}, U_3 = U_{3i}, U_4 = U_{4i}, U_5 = U_{5i}, U_6 = U_{6i}, \varepsilon_X = \varepsilon_{X_i}\right) \quad \text{(S8.1)}$$

$$\beta_{X_i} = f_{\beta_X}\left(Z = Z_i, U_1 = U_{1i}, U_4 = U_{4i}, U_6 = U_{6i}, \varepsilon_{\beta_X} = \varepsilon_{\beta_{X_i}}\right) \quad \text{(S9.1)}$$

$$\beta_{Y_i} = f_{\beta_Y}\left(X = X_i, U_2 = U_{2i}, U_5 = U_{5i}, U_6 = U_{6i}, \varepsilon_{\beta_Y} = \varepsilon_{\beta_{Y_i}}\right) \quad \text{(S10.1)}$$

In equations (S1.1)-(S6.1), $\varepsilon_L$ denotes the latent variable that is a cause of all unmeasured variables, thus rendering all of them d-connected to one another.

Since d-separation implies statistical independence, Figure 2 and equations S1.1-S10.1 can be used to find conditions under which $\beta_Y$ is d-separated to $\beta_X$ and $Z$, and thus statistically independent. From equations S9.1 and S10.1, it is clear that $\beta_X$ and $\beta_Y$ are d-connected due to shared terms (i.e., $U_6$), terms that may be correlated (e.g., $U_4$ and $U_5$) and terms that are functionally related (e.g., $X_i$ and $Z_i$) in the r.h.s. of these equations. Moreover, since equation S10.1 contains $X$ in the r.h.s., $\beta_Y$ and $Z$ are clearly d-connected.

Given Assumption 1, equations S1.1-S10.1 become:

$$U_{1_i} = f_{U_1}\left(Z = Z_i, \varepsilon_{L_1} = \varepsilon_{L_{1_i}}, \varepsilon_{U_1} = \varepsilon_{U_{1_i}}\right) \quad \text{(S1.2)}$$

$$U_{2_i} = f_{U_2}\left(\varepsilon_{L_2} = \varepsilon_{L_{2_i}}, \varepsilon_{U_2} = \varepsilon_{U_{2_i}}\right) \quad \text{(S2.2)}$$

$$U_{3_i} = f_{U_3}\left(\varepsilon_{L_1} = \varepsilon_{L_{1_i}}, \varepsilon_{L_2} = \varepsilon_{L_{2_i}}, \varepsilon_{U_3} = \varepsilon_{U_{3_i}}\right) \quad \text{(S3.2)}$$

$$U_{4_i} = f_{U_4}\left(\varepsilon_{L_1} = \varepsilon_{L_{1_i}}, \varepsilon_{U_4} = \varepsilon_{U_{4_i}}\right) \quad \text{(S4.2)}$$

$$U_{5_i} = f_{U_5}\left(\varepsilon_{L_2} = \varepsilon_{L_{2_i}}, \varepsilon_{U_5} = \varepsilon_{U_{5_i}}\right) \quad \text{(S5.2)}$$

$$U_{6_i} = \emptyset \quad \text{(S6.2)}$$

$$Z_i = f_Z\left(\varepsilon_Z = \varepsilon_{Z_i}\right) \quad \text{(S7.2)}$$

$$X_i = f_X\left(Z = Z_i, U_1 = U_{1_i}, U_3 = U_{3_i}, U_4 = U_{4_i}, U_5 = U_{5_i}, \varepsilon_X = \varepsilon_{X_i}\right) \quad \text{(S8.2)}$$

$$\beta_{X_i} = f_{\beta_X}\left(Z = Z_i, U_1 = U_{1_i}, U_4 = U_{4_i}, \varepsilon_{\beta_X} = \varepsilon_{\beta_{X_i}}\right) \quad \text{(S9.2)}$$

$$\beta_{Y_i} = f_{\beta_Y}\left(X = X_i, U_2 = U_{2_i}, U_5 = U_{5_i}, \varepsilon_{\beta_Y} = \varepsilon_{\beta_{Y_i}}\right) \quad \text{(S10.2),}$$

where $\varepsilon_{L_1}$ and $\varepsilon_{L_2}$ such that $\varepsilon_{L_1} \perp\!\!\!\perp \varepsilon_{L_2}$ are latent variables that allow $U_1$, $U_3$ and $U_4$ to be correlated with one another and $U_2$, $U_3$ and $U_5$ to be correlated with one another.

Assumption 2 implies that $X$ is not in the r.h.s. of equation S10.2, which becomes:

$$\beta_{Y_i} = f_{\beta_Y}\left(U_2 = U_{2_i}, U_5 = U_{5_i}, \varepsilon_{\beta_Y} = \varepsilon_{\beta_{Y_i}}\right) \quad \text{(S10.3).}$$

## 2. Implications of non-linear effects and data-generating models for NOSH

Under linear models for $X$ and $Y$, it is worth emphasizing the role of non-linear effects for NOSH. As can be seen in Figure 2A, if both the association of $Z$ on $X$ and the effect of $X$ on $Y$ are non-linear on the additive scale, the following path exists: $\beta_X \leftarrow Z \rightarrow X \rightarrow \beta_Y$, implying that NOSH is violated. That is, when both the association of $Z$ and $X$ and the effect of $X$ on $Y$ are non-linear, NOSH is violated regardless of unmeasured effect modifiers due to violations of Assumption 2.

We now discuss implications of multiplicative models. For simplicity of exposition, assume that $Z$, $X$, $Y$, $V_X$ and $V_Y$ (with $V_X$ and $V_Y$ denoting unmeasured variables) are all binary. $X_i \sim \text{Bernoulli}(\pi_{X_i})$, where $0 \leq \pi_{X_i} = \omega_X \times \alpha^{Z_i} \times \tau^{V_{X_i}} \leq 1$. This model can be re-written as $\ln(\text{E}[X|Z, V_X]) = \ln(\omega_X) +$ $Z\ln(\alpha) + V_X\ln(\theta)$, which clearly shows there is no multiplicative effect modification. This could be interpreted as $V_X \in U_3$. However, lack of multiplicative effect modification implies additive effect modification, which is the relevant scale for $\beta_X$. In this example, this happens because:

$$\mathrm{E}[X|Z = 1, V_X = 0] - \mathrm{E}[X|Z = 0, V_X = 0] = \omega_X\alpha - \omega_X = \omega_X(\alpha - 1)$$

$$\mathrm{E}[X|Z = 1, V_X = 1] - \mathrm{E}[X|Z = 0, V_X = 1] = \omega_X\alpha\tau - \omega_X\tau = \omega_X\tau(\alpha - 1)$$

The above illustrates that, under a multiplicative model on $X$, $\beta_X$ will likely vary according to other causes of $X$ (in this case, $V_X$), unless in rather contrived scenarios the model parameters cancel one another with respect to additive effect modification (e.g., if multiplicative effect modification leads exactly to no additive effect modification). Graphically, this implies that the causal diagram should contain an arrow from $V_X$ to $\beta_X$ under a multiplicative model on $X$. Therefore, $V_X \notin U_3$, being a potential member of $U_1$, $U_4$ or $U_6$.

Now suppose that $Y_i \sim \mathrm{Bernoulli}(\pi_{Y_i})$, where $0 \leq \pi_{Y_i} = \mu_Y \times \rho^{X_i} \times \chi^{V_{Y_i}} \leq 1$. Again, there is no multiplicative effect modification, but there is additive effect modification:

$$\mathrm{E}[Y|X = 1, V_Y = 0] - \mathrm{E}[Y|X = 0, V_Y = 0] = \mu_Y\rho - \mu_Y = \mu_Y(\rho - 1)$$

$$\mathrm{E}[Y|X = 1, V_Y = 1] - \mathrm{E}[Y|X = 0, V_Y = 1] = \mu_Y\rho\chi - \mu_Y\chi = \mu_Y\chi(\rho - 1)$$

The above illustrates that, under a multiplicative model on $Y$, $\beta_Y$ will likely vary according to other causes of $Y$ (in this case, $V_Y$). Graphically, this implies that the causal diagram should contain an arrow from $V_X$ to $\beta_X$. Therefore, $V_X \notin U_3$, being a potential member of $U_2$, $U_5$ or $U_6$.

This simple example illustrates the more general implication of the fact that additive effect modification is implied by lack of multiplicative effect modification for the plausibility of NOSH. Under a multiplicative model for $X$ and a linear model for $Y$, any cause of $X$ (here denoted by $V_X$) will generally be an effect modifier of $\beta_X$. Therefore, if $V_X$ is also either an effect modifier or a surrogate effect modifier of $\beta_Y$, Assumption 1 is violated. This implies that, unless heterogeneity in $\beta_Y$ is independent of all causes of $X$, Assumption 1 will likely be violated.

Under a linear model for $X$ and a multiplicative model for $Y$, any cause of $Y$ (here denoted by $V_Y$) will generally by an effect modifier of $\beta_Y$. Therefore, if $V_Y$ is also either an effect modifier or a surrogate effect modifier of $\beta_X$, then Assumption 1 would be violated. This implies that, unless heterogeneity in $\beta_X$ is independent of all causes of $Y$, Assumption 1 would likely be violated.

Under multiplicative models for both $X$ and $Y$, both $V_X$ and $V_Y$ will generally be modifiers of $\beta_X$ and $\beta_Y$, respectively. This implies that, if $V_X$ and $V_Y$ are statistically dependent, then both will likely be either effect modifiers or surrogate effect modifiers of both $\beta_X$ and $\beta_Y$, thus violating Assumption 1. This will happen, for example, if $V_X = V_Y$ or if these variables have a common cause (i.e., if there is unmeasured confounding). Since assuming no unmeasured confounding would eliminate the need of using instrumental variables, this is not a sensible assumption in this context.

## 3. Proof of theorem 1

We now prove Theorem 1. From the notation introduced in section 2.2 (recall that $U_1$ is not a cause of $Y$ and $U_3$ and $U_4$ do not modify the causal effect of $X$ on $Y$), the non-parametric structural equation model governing $Y$ can be equivalently defined as $Y_i = h(X_i, U_{2_i}, U_{5_i}, U_{6_i}, \varepsilon_{Y_i}) + g(U_{2_i}, U_{3_i}, U_{4_i}, U_{5_i}, U_{6_i}, \varepsilon_{Y_i})$, where $h(.)$ is a generic function such that that all of its terms include $X$, and $g(.)$ is another generic function that does not include $X$ as one of its arguments. From this, the function $F_{Y_i}(x)$ defined in section 2.1 can be equivalently defined as $F_{Y_i}(X_i) = \mathrm{E}\left[h(X_i, U_{2_i}, U_{5_i}, U_{6_i}, \varepsilon_{Y_i}) + g(U_{2_i}, U_{3_i}, U_{4_i}, U_{5_i}, U_{6_i}, \varepsilon_{Y_i}) | do(X_i = x), U = U_i, \varepsilon_Y = \varepsilon_{Y_i}\right]$.

From this, the numerator of the Wald estimand ($\mathrm{cov}(Y_i, Z_i)$) can defined as follows:

$$\mathrm{E}[Y_i Z_i] = \mathrm{E}\left[\left(h(X_i, U_{2_i}, U_{5_i}, U_{6_i}, \varepsilon_{Y_i}) + g(U_{2_i}, U_{3_i}, U_{4_i}, U_{5_i}, U_{6_i}, \varepsilon_{Y_i})\right) Z_i\right]$$

$$= \mathrm{E}\left[h(X_i, U_{2_i}, U_{5_i}, U_{6_i}, \varepsilon_{Y_i}) Z_i\right] + \mathrm{E}\left[g(U_{2_i}, U_{3_i}, U_{4_i}, U_{5_i}, U_{6_i}, \varepsilon_{Y_i})\right]\mathrm{E}[Z_i],$$

where $\mathrm{E}\left[g(U_{2_i}, U_{3_i}, U_{4_i}, U_{5_i}, U_{6_i}, \varepsilon_{Y_i}) Z_i\right] = \mathrm{E}\left[g(U_{2_i}, U_{3_i}, U_{4_i}, U_{5_i}, U_{6_i}, \varepsilon_{Y_i})\right]\mathrm{E}[Z_i]$ holds given the independence assumption (section 2.1).

$$\mathrm{E}[Y_i]\mathrm{E}[Z_i] = \mathrm{E}\left[h(X_i, U_{2_i}, U_{5_i}, U_{6_i}, \varepsilon_{Y_i}) + g(U_{2_i}, U_{3_i}, U_{4_i}, U_{5_i}, U_{6_i}, \varepsilon_{Y_i})\right]\mathrm{E}[Z_i]$$

$$= \mathrm{E}\left[h(X_i, U_{2_i}, U_{5_i}, U_{6_i}, \varepsilon_{Y_i})\right]\mathrm{E}[Z_i] + \mathrm{E}\left[h_2(U_{2_i}, U_{3_i}, U_{4_i}, U_{5_i}, U_{6_i}, \varepsilon_{Y_i})\right]\mathrm{E}[Z_i]$$

$$\mathrm{cov}(Y_i, Z_i) = \mathrm{E}[Y_i Z_i] - \mathrm{E}[Y_i]\mathrm{E}[Z_i]$$

$$= \mathrm{E}\left[h(X_i, U_{2_i}, U_{5_i}, U_{6_i}, \varepsilon_{Y_i}) Z_i\right] - \mathrm{E}\left[h(X_i, U_{2_i}, U_{5_i}, U_{6_i}, \varepsilon_{Y_i})\right]\mathrm{E}[Z_i]$$

$$= \mathrm{cov}\left(h(X_i, U_{2_i}, U_{5_i}, U_{6_i}, \varepsilon_{Y_i}), Z_i\right)$$

$$= \mathrm{cov}\left(\mathrm{E}\left[h(X_i, U_{2_i}, U_{5_i}, U_{6_i}, \varepsilon_{Y_i}) | X = X_i, U = U_i, \varepsilon_Y = \varepsilon_{Y_i}\right], Z_i\right).$$

The last equality follows from $\text{cov}(Y_i, Z_i) = \text{cov}\big(F_{Y_i}(X_i), Z_i\big)$ and that, for individual $i$, conditioning on $do(X = X_i)$ is equivalent to conditioning on the observed value of $X$ under the stable unit treatment value assumption.

Under Assumption 1, $U_6 = \emptyset$. Under Assumption 2, $\text{E}\big[h\big(X_i, U_{2_i}, U_{5_i}, U_{6_i}, \varepsilon_{Y_i}\big) | X = X_i, U = U_i, \varepsilon_Y = \varepsilon_{Y_i}\big] = \text{E}\big[h^{\#}\big(U_{2_i}, U_{5_i}, U_{6_i}, \varepsilon_{Y_i}\big) | U = U_i, \varepsilon_Y = \varepsilon_{Y_i}\big] X_i$, where $h^{\#}(.)$ is a generic function. In this case, $F_{Y_i}(1) - F_{Y_i}(0) = \text{E}\big[h^{\#}\big(U_{2_i}, U_{5_i}, U_{6_i}, \varepsilon_{Y_i}\big) | U = U_i, \varepsilon_Y = \varepsilon_{Y_i}\big]$ and (assuming differentiability with respect to $X$) $\frac{\partial}{\partial x}\big[F_{Y_i}(x)\big]\Big|_{x=X_i} = \text{E}\big[h^{\#}\big(U_{2_i}, U_{5_i}, U_{6_i}, \varepsilon_{Y_i}\big) | U = U_i, \varepsilon_Y = \varepsilon_{Y_i}\big]$. That is, for either a binary or a continuous $X$, $\beta_{Y_i} = \text{E}\big[h^{\#}\big(U_{2_i}, U_{5_i}, U_{6_i}, \varepsilon_{Y_i}\big) | U = U_i, \varepsilon_Y = \varepsilon_{Y_i}\big]$.

Therefore, if NOSH holds, $\text{cov}(Y_i, Z_i) = \text{cov}\big(\beta_{Y_i} X_i, Z_i\big) = \text{E}\big[\beta_{Y_i}\big]\text{cov}(X_i, Z_i)$. The last equality holds because:

$$\begin{aligned}\text{cov}\big(\beta_{Y_i} X_i, Z_i\big) &= \text{cov}\big(\text{E}\big[\beta_{Y_i} X_i | Z_i\big], \text{E}\big[Z_i | \beta_{Y_i}\big]\big) + \text{E}\big[\text{cov}\big(\beta_{Y_i} X_i, Z_i | \beta_{Y_i}\big)\big] \\ &= \text{E}\big[\text{cov}\big(\beta_{Y_i} X_i, Z_i | \beta_{Y_i}\big)\big].\end{aligned}$$

This equality holds because $\text{E}\big[Z_i | \beta_{Y_i}\big] = \text{E}[Z_i]$, so that $\text{cov}\big(\text{E}\big[\beta_{Y_i} X_i | Z_i\big], \text{E}\big[Z_i | \beta_{Y_i}\big]\big) = \text{cov}\big(\text{E}\big[\beta_{Y_i} X_i | Z_i\big], \text{E}[Z_i]\big) = 0$. Therefore:

$$\begin{aligned}\text{cov}\big(\beta_{Y_i} X_i, Z_i\big) &= \text{E}\big[\text{cov}\big(\beta_{Y_i} X_i, Z_i | \beta_{Y_i}\big)\big] = \text{E}\big[\beta_{Y_i}\text{cov}\big(X_i, Z_i | \beta_{Y_i}\big)\big] \\ &= \text{E}\big[\beta_{Y_i}\text{cov}(X_i, Z_i)\big] = \text{E}\big[\beta_{Y_i}\big]\text{cov}(X_i, Z_i).\end{aligned}$$

The equality $\text{E}\big[\beta_{Y_i}\text{cov}\big(X_i, Z_i | \beta_{Y_i}\big)\big] = \text{E}\big[\beta_{Y_i}\text{cov}(X_i, Z_i)\big]$ holds because, by NOSH, $\beta_Y \perp\!\!\!\perp \beta_X$.

From this, $\beta_{IV} = \frac{\text{E}[\beta_{Y_i}]\text{cov}(X_i, Z_i)}{\text{cov}(X_i, Z_i)} = \text{E}\big[\beta_{Y_i}\big] = \text{ACE}$, thus finishing the proof.

Although the above proves Theorem 1 in generality, it is instructive to consider alternative proofs under specific scenarios. We will also use these additional proofs to show that only mean independence is sufficient for identification. Initially, consider the case where $Z$ and $X$ are binary. In this case, $\beta_{X_i} \in \{1, -1, 0\}$, respectively indicating that individual $i$ is a complier, a defier or either a never or always taker. In this case, $\beta_{IV}$ can be defined as $\beta_{IV} = \frac{\text{E}[\beta_{Y_i} | \beta_{X_i} = 1] P(\beta_{X_i} = 1) - \text{E}[\beta_{Y_i} | \beta_{X_i} = -1] P(\beta_{X_i} = -1)}{P(\beta_{X_i} = 1) - P(\beta_{X_i} = -1)}$, which is essentially a weighted average of conditional ACEs within subgroups where the relevance assumption hold, with the sign of the weights corresponding to the sign of the $Z$-$X$ association.[1] Because $X$ is binary, the effect of X on $Y$ is necessarily additive linear (i.e., Assumption 2 necessarily holds). If Assumption 1 also holds (even if $\beta_Y$ and $\beta_X$ are mean independent, but not fully independent), then $\mathrm{E}\left[\beta_{Y_i}|\beta_{X_i}=1\right]=\mathrm{E}\left[\beta_{Y_i}|\beta_{X_i}=-1\right]=\mathrm{E}\left[\beta_{Y_i}\right]=\mathrm{ACE}$. Since $\beta_{IV}$ is a weighted average of conditional effects, $\beta_{IV}=\mathrm{ACE}$.

Now, consider the case where $Z$ is binary and $X$ is continuous. In this case, the effects of $Z$ on $X$ and on $Y$ are necessarily additive linear, implying that $\beta_{X_i}=X_i(Z_i=1)-X_i(Z_i=0)=\frac{\partial X_i}{\partial Z}$, where $X_i(Z_i=z)=f_X(do(Z=z),U=U_i,\varepsilon_X=\varepsilon_{X_i})$ for $z\in\{0,1\}$. Therefore, the denominator of the Wald estimand is $\mathrm{E}\left[\beta_{X_i}\right]=\mathrm{E}\left[\frac{\partial X_i}{\partial Z}\right]$. The numerator of the Wald estimand is: $\mathrm{E}\left[Y_i(X_i(Z_i=1))-Y_i(X_i(Z_i=0))\right]=\mathrm{E}\left[\frac{\left(Y_i(X_i(Z_i=1))-Y_i(X_i(Z_i=0))\right)}{X_i(Z_i=1)-X_i(Z_i=0)}\left(X_i(Z_i=1)-X_i(Z_i=0)\right)\right]=$ $\mathrm{E}\left[\frac{\left(Y_i(X_i(Z_i=1))-Y_i(X_i(Z_i=0))\right)}{X_i(Z_i=1)-X_i(Z_i=0)}\frac{\partial X_i}{\partial Z}\right]$, where $Y_i(X_i(Z_i=z))=f_Y(do(X=X_i(Z_i=z)),U=U_i,\varepsilon_Y=\varepsilon_{Y_i})$. Notice that, for simplicity, the definitions of $X_i(Z_i=z)$ and $Y_i(X_i(Z_i=z))$ assume deterministic counterfactuals. Therefore, the numerator of the Wald estimand equals $\mathrm{E}\left[\frac{\partial Y_i}{\partial X}\frac{\partial X_i}{\partial Z}\right]$ if the effect of $X$ on $Y$ is additive linear, which is true under NOSH, which also implies $\frac{\partial Y_i}{\partial X}\perp\!\!\!\perp\frac{\partial X_i}{\partial Z}$ (or just $\mathrm{cov}\left(\frac{\partial Y_i}{\partial X},\frac{\partial X_i}{\partial Z}\right)=0$ under the relaxed version of NOSH). In this case, $\beta_{IV}=\frac{\mathrm{E}\left[\frac{\partial Y_i \partial X_i}{\partial X\,\partial Z}\right]}{\mathrm{E}\left[\frac{\partial X_i}{\partial Z}\right]}=\frac{\mathrm{E}\left[\frac{\partial Y_i}{\partial X}\right]\mathrm{E}\left[\frac{\partial X_i}{\partial Z}\right]}{\mathrm{E}\left[\frac{\partial X_i}{\partial Z}\right]}=\mathrm{E}\left[\frac{\partial Y_i}{\partial X}\right]=\mathrm{E}\left[\beta_{Y_i}\right]$.[2]

The case where $Z$ is multivalued follows from the arguments above. For a multivalued discrete $Z$ coded such that $\mathrm{E}[X|Z=1]\leq\cdots\leq\mathrm{E}[X|Z=K]$, where $K$ is the number of values that $Z$ attains, let $\beta_{IV_{z,z-1}}=\frac{\mathrm{E}[Y_i|Z=z]-\mathrm{E}[Y_i|Z=z-1]}{\mathrm{E}[X_i|Z=z]-\mathrm{E}[X_i|Z=z-1]}$. The IV estimand is $\beta_{IV}=\sum_{z=2}^{K}\beta_{IV_{z,z-1}}\omega_z$, where $\omega_z$ denotes the weight that $\beta_{IV_{z,z-1}}$ receives in the overall estimand, defined such that $\sum_{z=2}^{K}\omega_z=1$.[3,4] Notice that $\beta_{IV_{z,z-1}}$ is essentially the Wald estimand for a binary instrument but restricted to the subset $Z\in\{z-1,z\}$. Therefore, by the arguments above, $\beta_{IV_{z,z-1}}=\mathrm{E}\left[\beta_{Y_i}|Z_i\in\{z-1,z\}\right]=\mathrm{E}\left[\beta_{Y_i}\right]=\mathrm{ACE}$, where the second equality follows from the fact that, under NOSH, $\beta_Y\perp\!\!\!\perp(\beta_X,Z)$ (or just $\mathrm{E}[\beta_Y|Z,\beta_X]=\mathrm{E}[\beta_Y]$ under the relaxed version). This implies that the IV estimand is $\beta_{IV}=\sum_{z=2}^{K}\beta_{IV_{z,z-1}}\omega_z=\sum_{z=2}^{K}\mathrm{E}\left[\beta_{Y_i}\right]\omega_z=\mathrm{E}\left[\beta_{Y_i}\right]$. The case for a continuous $Z$ is similar, where summation over $Z$ is replaced with integration over $Z$: that is, $\beta_{IV}=\int_{-\infty}^{\infty}\beta_{IV_z}\varphi_z dz$, where $\beta_{IV_z}=\lim_{d\to 0^+}\frac{\mathrm{E}[Y_i|Z=z+d]-\mathrm{E}[Y_i|Z=z]}{\mathrm{E}[X_i|Z=z+d]-\mathrm{E}[X_i|Z=z]}$ and $\varphi_z$ are weights defined such that $\int_{-\infty}^{\infty}\varphi_z dz=1$.[3] Under NOSH, $\beta_{IV_z}=\lim_{d\to 0^+}\mathrm{E}\left[\beta_{Y_i}|Z\{z,z+d\}\right]=\lim_{d\to 0^+}\mathrm{E}\left[\beta_{Y_i}\right]=\mathrm{E}\left[\beta_{Y_i}\right]$. Therefore, $\beta_{IV}=\int_{-\infty}^{\infty}\beta_{IV_z}\varphi_z dz=\int_{-\infty}^{\infty}\mathrm{E}\left[\beta_{Y_i}\right]\varphi_z dz=\mathrm{E}\left[\beta_{Y_i}\right]$.

# 4. Simulation study

## 4.1. Data-generating model

We performed simulations to further demonstrate that IVs identify the ACE under NOSH, evaluating several related examples. We used the following data-generating model where $Z$, $X$ and $Y$ are continuous:

$$Z_i = \varepsilon_{Z_i} \quad \text{(S11)},$$

$U_{k_i} \sim \text{Binom}(0.5)$ for $k \in \{3,4,5,6\}$, independently (S12),

$V_{k_i} \sim \text{Binom}(0.5)$ for $k \in \{3,4,5,6\}$, independently (S13),

$$X_i \sim \gamma Z_i + \rho Z_i^3 + \sum_{k=3}^{6} \left( \delta_{X_k}^{U} U_{k_i} + \delta_{X_k}^{V} V_{k_i} + \delta_{X_k}^{U\times V} U_{k_i} V_{k_i} \right) + \sum_{k\in\{4,6\}} \left( \theta_{X_k}^{U} Z_i U_{k_i} + \theta_{X_k}^{V} Z_i V_{k_i} + \theta_{X_k}^{U\times V} Z_i U_{k_i} V_{k_i} \right) + \varepsilon_{X_i} \quad \text{(S14)},$$

$$Y_i \sim \tau X_i + \varphi X_i^2 + \sum_{k=3}^{6} \left( \delta_{Y_k}^{U} U_{k_i} + \delta_{Y_k}^{V} V_{k_i} + \delta_{Y_k}^{U\times V} U_{k_i} V_{k_i} \right) + \sum_{k\in\{5,6\}} \left( \theta_{Y_k}^{U} X_i U_{k_i} + \theta_{Y_k}^{V} X_i V_{k_i} + \theta_{Y_k}^{U\times V} X_i U_{k_i} V_{k_i} \right) + \varepsilon_{Y_i} \quad \text{(S15)},$$

where $\varepsilon_{Z_i}, \varepsilon_{X_i}, \varepsilon_{Y_i}$ are independent, distributed as described below. $k$ starts at 3 instead of 1 so that the classification in Table 1 also applies here.

$\{U_k\}_{k=3}^{6}$ and $\{V_k\}_{k=3}^{6}$ are two sets of independent binary confounders between the $X$ and the $Y$. For all $k \in \{3,4,5,6\}$, $P(U_k = 1) = P(V_k = 1) = 0.5$. From the model, the confounders can be classified as in Table 1, as follows:

- $U_3$ and $V_3$ do not modify either the $Z$-$X$ effect or the $X$-$Y$ effect.
- $U_4$ and $V_4$ modify the effect of $Z$ on $X$ (parameters $\theta_{X_4}^{U}$, $\theta_{X_4}^{V}$ and $\theta_{X_4}^{U\times V}$ in (S14)), but not the effect of $X$ on $Y$.
- $U_5$ and $V_5$ modify the effect of effect of $X$ on $Y$ (parameters $\theta_{Y_5}^{U}$, $\theta_{Y_5}^{V}$ and $\theta_{Y_5}^{U\times V}$ in (S15)), but not the $Z$-$X$ effect.

- $U_6$ and $V_6$ modify the effect of $Z$ on $X$ (parameters $\theta_{X_6}^{U}$, $\theta_{X_6}^{V}$ and $\theta_{X_6}^{U\times V}$ in (S14)), and the effect of $X$ on $Y$ (parameters $\theta_{Y_4}^{U}$, $\theta_{Y_4}^{V}$ and $\theta_{Y_4}^{U\times V}$ in (S15)).

The data-generating model also allows the $Z$-$X$ and $X$-$Y$ effects to be non-linear if $\rho \neq 0$ and $\varphi \neq 0$, respectively.

### 4.2. Simulation scenarios

In all simulations, $\delta_{X_k}^{U} = \delta_{X_k}^{V} = \delta_{X_k}^{U\times V} = \delta_{Y_k}^{U} = \delta_{Y_k}^{V} = \delta_{Y_k}^{U\times V} = 0.5$ for all $k \in \{3,4,5,6\}$, implying that the unconditional ordinary lest squares estimate is upwardly biased for the causal effect of $X$ on $Y$. Effect heterogeneity can occur on the effect of $Z$ on $X$ and/or on the effect of $X$ on $Y$.

#### 4.2.1. Scenario 1: NOSH holds

In scenario 1, data was generated under NOSH by setting $\theta_{X_6}^{U} = \theta_{X_6}^{V} = \theta_{X_6}^{U\times V} = \theta_{Y_6}^{U} = \theta_{Y_6}^{V} = \theta_{Y_6}^{U\times V} = \varphi = 0$. Moreover, we set $\tau = 0.5$, $\theta_{Y_5}^{U} = 0.5$, $\theta_{Y_5}^{V} = -1$ and $\theta_{Y_5}^{U\times V} = 0$, so that the effect of $X$ on $Y$ is linear within strata of $U_5$ and $V_5$. Since both $U_5$ and $V_5$ affect $X$, NEM1 and NEM2 are violated. The effect of $Z$ on $X$ was regulated by setting $\gamma = -0.4$, $\rho = 0.3$, $\theta_{X_4}^{U} = 0.5$, $\theta_{X_4}^{V} = -1$ and $\theta_{X_4}^{U\times V} = 0$.

#### 4.2.2. Scenarios 2 and 3: NOSH is violated

In scenario 2, assumption 1 (but not assumption 2) is violated. This was done by setting $\theta_{X_k}^{U} = \theta_{X_k}^{V} = \theta_{X_k}^{U\times V} = \theta_{Y_k}^{U} = \theta_{Y_k}^{V} = \theta_{Y_k}^{U\times V} = \varphi = 0$ for $k \in \{4,5\}$. The causal effect of $X$ on $Y$ is linear within strata of $U_6$ and $V_6$, and was regulated by setting $\tau = \theta_{Y_6}^{U} = \theta_{Y_6}^{V} = \theta_{Y_6}^{U\times V} = 0.5$. The effect of $Z$ on $X$ was regulated by setting $\gamma = -0.4$, $\rho = 0.3$ and $\theta_{X_6}^{U} = \theta_{X_6}^{V} = \theta_{X_6}^{U\times V} = 0.5$. In scenario 3, assumption 2 (but not assumption 1) is violated. This was done by setting the parameters in the same way as in scenario 1, except that $\tau = 0$ and $\varphi = 0.2$.

#### 4.2.3. Scenarios 4 and 5: non-normal error terms

In scenarios 1-3, $\varepsilon_{Z_i}, \varepsilon_{X_i}, \varepsilon_{Y_i} \sim N(0,1)$, independently. To corroborate the notion that NOSH does not require normal error terms, scenario 1 was re-evaluated with the difference that error terms are independently sampled from the following distributions: $\mathrm{Beta}(0.5,0.5)$ (scenario 4), $\chi^2(2) + 7I$, where $I \sim \mathrm{Bernoulli}(0.5)$ (scenario 5). In these scenarios, the error terms were converted to sample Z scores so that all have sample mean of 0 and sample variance of 1, to improve comparisons with scenario 1.

### 4.3. Statistical analysis

We used the two stage least squares estimator (TSLS) to estimate the $\mathrm{ACE}$. For the case of a single $Z$, $X$ and $Y$, TSLS is equivalent to the Wald estimator.[3,5]. This corresponds to assuming that NOSH holds marginally. We refer to such TSLS specification – i.e., a model with no covariates – as TSLS(1). We also considered three additional TSLS specifications, where $U_6$ and $V_6$ are incorporated in the model in different ways. To simplify the interpretation of the model, $U_6$ and $V_6$ were converted to deviations from their sample means.

- TSLS(2): including $U_6$ and $V_6$ as covariates, with no product term. That is:

$$\hat{X}_i = \widehat{\beta_0^X} + \widehat{\beta_1^X} Z_i + \widehat{\Psi_U^X} U_{6_i} + \widehat{\Psi_V^X} V_{6_i}$$

$$\hat{Y}_i = \widehat{\beta_0^Y} + \widehat{\beta_1^Y} \hat{X}_i + \widehat{\Psi_U^Y} U_{6_i} + \widehat{\Psi_V^Y} V_{6_i}$$

- TSLS(3): including $U_6$ and $V_6$, but not $U_6 \times V_6$, as covariates, with product terms with $Z$ and $X$. That is:

$$\hat{X}_i = \widehat{\beta_0^X} + \widehat{\beta_1^X} Z_i + \widehat{\Psi_U^X} U_{6_i} + \widehat{\Psi_V^X} V_{6_i} + \widehat{\lambda_U^X} U_{6_i} Z_i + \widehat{\lambda_V^X} V_{6_i} Z_i$$

$$\hat{Y}_i = \widehat{\beta_0^Y} + \widehat{\beta_1^Y} \hat{X}_i + \widehat{\Psi_U^Y} U_{6_i} + \widehat{\Psi_V^Y} V_{6_i} + \widehat{\lambda_U^Y} U_{6_i} Z_i + \widehat{\lambda_V^Y} V_{6_i} Z_i$$

- TSLS(4): including $U_6$, $V_6$ and $U_6 \times V_6$ as covariates. That is:

$$\hat{X}_i = \widehat{\beta_0^X} + \widehat{\beta_1^X} Z_i + \widehat{\Psi_U^X} U_{6_i} + \widehat{\Psi_V^X} V_{6_i} + \widehat{\Psi_{U\times V}^X} V_{6_i} + \widehat{\lambda_U^X} U_{6_i} Z_i + \widehat{\lambda_V^X} V_{6_i} Z_i + \widehat{\lambda_{U\times V}^X} U_{6_i} V_{6_i} Z_i$$

$$\hat{Y}_i = \widehat{\beta_0^Y} + \widehat{\beta_1^Y} \hat{X}_i + \widehat{\Psi_U^Y} U_{6_i} + \widehat{\Psi_V^Y} V_{6_i} + \widehat{\Psi_{U\times V}^Y} V_{6_i} + \widehat{\lambda_U^Y} U_{6_i} Z_i + \widehat{\lambda_V^Y} V_{6_i} Z_i + \widehat{\lambda_{U\times V}^Y} U_{6_i} V_{6_i} Z_i$$

Assuming NOSH holds given $U_6$ and $V_6$ (as in scenario 2 of the simulation study), consistency of $\widehat{\beta_1^Y}$ as an estimate of the $\mathrm{ACE}$ also requires correct model specification and that covariates were measured without error. Interacting each covariate with $Z$ and $X$ in this way improves robustness to model specification.[6] In practice, it may be useful to compare different specifications to assess sensitivity to model misspecification.

We calculated median bias and standard error, coverage (here defined as the proportion of times that the 95% confidence intervals included the $\mathrm{ACE}$) and rejection rate (here defined as the proportion of times that the 95% confidence intervals excluded the null) across 20,000 simulated datasets. Confidence intervals were calculated using Huber-White standard errors for instrumental variable analysis.[7]

## 5. Re-examination of selected published studies

NOSH is an untestable assumption that cannot be guaranteed by study design, and thus assessing its plausibility requires subject matter knowledge. Indeed, in some applications this assumption may be implausible. We now discuss three published IV studies to further illustrate how NOSH can be used for interpreting the Wald estimate as a consistent estimate for the ACE.

### 5.1. Treatment allocation in a RCT of vitamin A on mortality

Sommer et al. (1986)[8] ran a RCT investigating the effect of vitamin A supplementation (which corresponds to the treatment $X$) on childhood mortality (which corresponds to the treatment $Y$) in 450 villages in northern Sumatra (Indonesia). About half of the villages were allocated to each group at random (which corresponds to the instrument $Z$). 93.2% of those allocated to treatment took at least one vitamin A tablet, compared to only 1.1% of those allocated to control. The estimated effect of $Z$ on $X$ was thus 92.1 percent points (95% confidence interval [CI]: 91.6 to 92.7) per 100. The estimated effect of the intervention allocation on mortality was 0.29 (95% CI: 0.05 to 0.52). The Wald estimate of the effect of taking at least one pill on mortality was therefore 0.31 (95% CI: 0.05 to 0.57).

Under monotonicity, the IV estimate could be interpreted as the effect of taking at least one vitamin A tablet amongst those who would always comply to the assigned treatment. Such subgroup of the population is therefore often referred as compliers. For binary $Z$ (in the example, $Z = 0$ and $Z = 1$ respectively denote being allocated to vitamin A or control) and $X$ (in the example, $X = 0$ and $X = 1$ respectively denote actually taking vitamin A or not), monotonicity is defined as $X_i(Z_i = 1) \geq X_i(Z_i = 0)$ for everyone in the population, with $X_i(Z_i = z)$ defined in the same way as in section 3 in the Supplement. Individual $i$ is a complier if $X_i(Z_i = z) = z \in \{0,1\}$.

Monotonicity is frequently proposed as an assumption that allows interpreting the IV estimate as consistent for a well-defined causal estimand (the ACE among compliers) when neither NEM nor instrument effect homogeneity hold, which are indeed often strong assumptions. However, under the assumption that the individual level effects of $Z$ on $X$ and of $X$ on $Y$ are independent, which is weaker than other IV4 assumptions known to identify the ACE (as described in section 2.3), NOSH holds (because Assumption 1 is satisfied and, because $X$ is binary, Assumption 2 is automatically satisfied) the IV estimate can be interpreted as a consistent estimate of the ACE in the population even if other IV4 assumptions are violated. Of note, even if monotonicity is violated, the IV estimate can still be interpreted as an estimate of the ACE under NOSH (or any of the other IV4 assumptions).

In this example, NOSH requires that the participants' compliance behavior was independent of their potential outcomes. Given *a priori*, it is plausible to assume that the effects of taking vitamin A were generally not known by the study participants. However, this is an untestable assumption. A plausible situation where NOSH would be violated is if compliance were lower and the effect of vitamin A were greater among study participants from disadvantaged villages. In this case, weaker treatment effects would receive greater weight than stronger treatment effects (because the first occurs more often in subgroups of the study sample where the IV is stronger), thus leading to bias in the IV estimate if interpreted as an estimate of the ACE across the population. The magnitude of this bias will be proportional to the correlation between the individual-level effects of $Z$ on $X$ and of $X$ on $Y$.

### 5.2. Using draft lottery to estimate the effect of veteran status on earnings

Angrist (1990)[9] investigated the effect of serving in the military ($X$) on taxable earnings ($Y$). He used the Vietnam draft lottery ($Z$) as an instrument for veteran status. Men were selected for the draft by randomly dates of birth within each year. Men whose birthday was selected were eligible for the draft; all other men could volunteer but were not drafted. After the lottery, eligible men were screened for physical and mental criteria, and some were eliminated after these screens. Men who were eligible for the draft because of the lottery were between 10 and 16 percentage points more likely to be a veteran. Eligible men had lower earnings. The Wald estimate of the effect of serving in the military on earnings was -$1920 (95%CI: -$3049 to -$792) per year in 1978 dollars.

In this example, again Assumption 2 is automatically satisfied because $X$ is binary. NOSH would therefore require that the individual-level effects of the lottery draft on the probability of being a veteran are independent of the individual-level causal effects of veteran status on earnings. NOSH could be violated, for example, if men who were drafted, but did not become veterans, had particularly large differences in potential earnings – i.e., the causal effect is greater among those who did not serve in the military even though they were drafted (i.e., never takers and/or defiers). In case such correlation between the individual-level effects of the lottery on veteran status and the individual-level effects of serving on earnings exists, then the Wald estimate would not be consistent for the $\mathrm{ACE}$.

### 5.3. Mendelian randomization: the effect of body mass index on coronary heart disease

As a final illustration, we discuss a Mendelian randomization study, where genetic variants robustly associated with the treatment variable are used as IVs.[4,10] Here, we discuss a study by Dale et al. (2017)[11] investigating the effects of body mass index (BMI) on coronary heart disease (CHD). They measured 97 genetic variants robustly associated with BMI in 14 prospective studies and used these genetic variants as IVs. They combined the 97 variants into a single genetic score and found that 1 standard deviation increase in BMI increased odds of CHD by 36% (95% CI: 22% to 52%).

The genetic score is likely to have heterogenous effects on BMI across the population,[12] but it may affect everyone's BMI in the same direction. In this case, the IV estimate can be interpreted as a weighted average of the effect of increasing BMI in all those individuals whose BMI was affected by the genetic score. Under monotonicity, each individual contributes to the IV estimate proportionally to the effect of the genetic score on their BMI.[4,13] Although well-defined mathematically, this estimand is difficult to interpret from a policy-making perspective, because different unknown subgroups of the population contribute different unknown weights.

In this example, it may me implausible that NOSH holds due to the evidence supporting non-linear associations between BMI and CHD-related outcomes,[14,15] which would violate Assumption 2. Assumption 1 relates to individual-level effects of the genetic score on BMI being independent of the individual-level effects of BMI on CHD. Is this plausible? Modifiers of the effect of the genetic score on BMI may correlate with modifiers of the effect of BMI on CHD. Indeed, results from the UK Biobank indicate that the effect of the genetic score on BMI varies according to many variables, such as alcohol intake and Townsend Deprivation Index (a measure of socioeconomic position).[16] Assumption 1 would be violated if the effect of BMI on CHD varies by one of these variables.

## 7. Supplementary Table

**Supplementary Table 1. Median bias, coverage and power of four two-stage least squares (TSLS) regression specifications as estimators of the ACE† in scenarios 1-5‡.**

| N | TSLS | Scenario 1 (ACE=0.250) | | | Scenario 2 (ACE=1.125) | | | Scenario 3 (ACE=0.750) | | | Scenario 4 (ACE=0.250) | | | Scenario 5 (ACE=0.250) | | |
|---|---|---|---|---|---|---|---|---|---|---|---|---|---|---|---|---|
| | | Bias | Coverage | Power | Bias | Coverage | Power | Bias | Coverage | Power | Bias | Coverage | Power | Bias | Coverage | Power |
| 250 | 1 | -0.001 | 95.4 | 19.4 | 1.262 | 75.3 | 63.4 | 0.407 | 65.8 | 94.3 | 0.001 | 95.6 | 23.1 | 0.003 | 96.7 | 19.6 |
| | 2 | -0.002 | 95.1 | 20.0 | 1.261 | 61.0 | 75.1 | 0.407 | 65.2 | 95.4 | 0.001 | 95.5 | 23.3 | 0.003 | 96.4 | 19.7 |
| | 3 | -0.010 | 96.3 | 15.9 | 0.144 | 95.2 | 60.9 | 0.380 | 70.9 | 87.8 | -0.012 | 96.4 | 19.2 | -0.010 | 97.7 | 14.3 |
| | 4 | -0.017 | 97.1 | 13.6 | 0.112 | 99.2 | 35.3 | 0.367 | 74.8 | 80.8 | -0.018 | 96.9 | 16.7 | -0.016 | 98.2 | 11.7 |
| 1000 | 1 | 0.002 | 95.0 | 53.8 | 1.316 | 14.3 | 98.8 | 0.414 | 18.2 | 100.0 | 0.001 | 95.3 | 62.1 | 0.000 | 95.4 | 51.8 |
| | 2 | 0.002 | 94.7 | 55.7 | 1.314 | 3.4 | 99.2 | 0.414 | 16.3 | 100.0 | 0.001 | 95.3 | 64.4 | 0.001 | 95.5 | 53.8 |
| | 3 | -0.001 | 94.8 | 53.1 | 0.143 | 84.1 | 91.0 | 0.407 | 18.1 | 100.0 | -0.002 | 95.4 | 61.7 | -0.003 | 95.7 | 50.2 |
| | 4 | -0.002 | 94.9 | 51.8 | 0.055 | 99.2 | 49.8 | 0.404 | 19.4 | 100.0 | -0.004 | 95.5 | 60.4 | -0.004 | 96.0 | 48.1 |
| 2500 | 1 | 0.000 | 95.1 | 88.1 | 1.319 | 0.1 | 100.0 | 0.413 | 0.6 | 100.0 | 0.000 | 95.0 | 93.6 | 0.000 | 95.1 | 86.7 |
| | 2 | -0.001 | 95.1 | 89.5 | 1.319 | 0.0 | 100.0 | 0.414 | 0.4 | 100.0 | 0.001 | 94.9 | 95.1 | 0.000 | 95.1 | 88.8 |
| | 3 | -0.002 | 95.0 | 88.7 | 0.146 | 64.7 | 98.5 | 0.412 | 0.5 | 100.0 | -0.001 | 95.0 | 94.6 | -0.001 | 95.3 | 87.5 |
| | 4 | -0.002 | 94.9 | 88.4 | 0.008 | 98.7 | 66.9 | 0.410 | 0.6 | 100.0 | -0.001 | 95.1 | 94.5 | -0.002 | 95.4 | 86.9 |
| 5000 | 1 | 0.000 | 95.0 | 99.2 | 1.321 | 0.0 | 100.0 | 0.417 | 0.0 | 100.0 | 0.000 | 95.0 | 99.7 | 0.000 | 95.2 | 98.9 |
| | 2 | 0.000 | 95.0 | 99.4 | 1.322 | 0.0 | 100.0 | 0.417 | 0.0 | 100.0 | 0.000 | 95.0 | 99.9 | 0.000 | 95.2 | 99.3 |
| | 3 | -0.001 | 95.0 | 99.4 | 0.145 | 40.3 | 99.9 | 0.415 | 0.0 | 100.0 | -0.001 | 95.1 | 99.8 | 0.000 | 95.2 | 99.2 |
| | 4 | -0.001 | 95.0 | 99.4 | -0.010 | 98.0 | 84.0 | 0.415 | 0.0 | 100.0 | -0.001 | 95.0 | 99.9 | -0.001 | 95.3 | 99.2 |
| 10000 | 1 | 0.000 | 94.8 | 100.0 | 1.320 | 0.0 | 100.0 | 0.418 | 0.0 | 100.0 | 0.001 | 94.9 | 100.0 | 0.000 | 95.3 | 100.0 |
| | 2 | 0.000 | 94.8 | 100.0 | 1.319 | 0.0 | 100.0 | 0.418 | 0.0 | 100.0 | 0.001 | 94.9 | 100.0 | 0.000 | 95.3 | 100.0 |
| | 3 | -0.001 | 94.8 | 100.0 | 0.146 | 14.1 | 100.0 | 0.416 | 0.0 | 100.0 | 0.000 | 94.9 | 100.0 | 0.000 | 95.3 | 100.0 |
| | 4 | -0.001 | 94.8 | 100.0 | -0.011 | 97.4 | 96.4 | 0.416 | 0.0 | 100.0 | 0.000 | 95.0 | 100.0 | 0.000 | 95.3 | 100.0 |

†Average causal effect (ACE) of a unit increase in $X$ on $Y$.
‡1: NOSH holds. 2: Assumption 1 violated. 3: Assumption 2 violated. 4: NOSH holds and error terms are drawn from a beta distribution. 5: NOSH holds and error terms are drawn from a mixed chi-squared distribution.